\documentclass[prb,twocolumn,showpacs,showkeys,preprintnumbers,amsmath,amssymb]{revtex4}

\usepackage[dvips]{graphicx}
\usepackage{bm}
\usepackage{epsfig}
\usepackage{subfigure}
\usepackage{dcolumn}
\usepackage{multirow}

\begin{document}

\preprint{accepted at Phys.\ Rev.\ B}

\title{
Exact-exchange based quasiparticle energy calculations for the band gap, 
effective masses and deformation potentials of ScN}
\author{Abdallah Qteish}
%\email{aqteish@yu.edu.jo} 
\affiliation{Department of Physics, Yarmouk University, 21163-Irbid, Jordan}
\author{Patrick Rinke and Matthias Scheffler}
\affiliation{Fritz-Haber-Institut der Max-Planck-Gesellschaft,
Faradayweg 4--6, D-14195 Berlin-Dahlem, Germany}
\author{J\"org Neugebauer}
\affiliation{MPI f\"ur Eisenforschung, Max-Planck Stra{\ss}e 1, D-40237 
D\"usseldorf, Germany}

\date{\today}

\begin{abstract}

The band gaps, longitudinal and transverse effective masses, and deformation 
potentials of ScN in the rock-salt structure have been calculated employing G$_0$W$_0$-quasiparticle
calculations using exact-exchange Kohn-Sham density functional theory one-particle 
wavefunctions and energies as input. 
Our quasiparticle gaps support recent experimental 
observations that ScN has a much lower indirect band gap than previously thought.
The results are analyzed in terms of the influence of 
different approximations for exchange and correlation taken in the
computational approach on the electronic structure of ScN.
\end{abstract}

\maketitle

\section{Introduction}

Scandium nitride (ScN) is emerging as a versatile material for promising technological
applications. As part of the transition metal nitride family it initially
generated interest for potential applications as wear resistant 
and optical coatings due to its 
mechanical strength, high melting point of ~2600$^\circ$ C \cite{Bai},
and high hardness ($H$ = 21 GPa) 
with respect to load deformations \cite{Gall98}.
ScN crystallizes in the rock-salt phase with a lattice parameter of 
4.50 \AA \cite{Dismukes72}. The octahedral bonding arrangement provides a much
more favorable environment for the
incorporation of transition metal atoms like Mn or Cr than the
tetrahedrally coordinated III-V semiconductors, which have up until now 
been popular candidates for spintronic materials. 
Successful incorporation of Mn into ScN
has been demonstrated \cite{Brithen04} and {\it ab initio} calculations predict
Mn-doped ScN to be a dilute ferromagnetic semiconductor \cite{Herwadkar05}.
Moreover, ScN has a lattice mismatch of less than 2\% to cubic gallium nitride
(GaN). This makes ScN structurally compatible with the group-IIIA nitrides
\cite{ScGaN_exp96,ScGaN_exp01,ScN/GaN_exp01,ScGaN/ScInN_th02,ScGaN_exp04,
      ScGaN_exp05,ScGaN/ScInN_th05_1,ScGaN/ScInN_th05_2}
 -- an important technological material class, in particular for applications 
in optoelectronic devices. 
Alloying ScN with GaN \cite{ScGaN_exp96,ScGaN_exp01,ScGaN_exp04,ScGaN_exp05}
might provide a viable alternative to InGaN alloys for use in light emitting
devices or solar cells. In addition multifunctional devices are conceivable if
the strong electromechanical response predicted for hexagonal ScN 
\cite{h-ScN_th05} can be utilized.

The electronic band structure of ScN -- a key quantity for the design of
optoelectronic devices -- has been difficult to access 
both experimentally and theoretically. Early experiments were 
hampered by various complications in growing films with well defined 
crystalline orientation, stoichiometry, low background carrier 
concentration, and surface roughness. For a detailed discussion we refer to 
e.g. Ref. \onlinecite{Gall01}. Recent advances in growth techniques 
%employing ultra-high vacuum (UHV) deposition systems 
have led to 
a systematic improvement of the material's quality \cite{Al-Brithen00}. 
Employing optical spectroscopy and photoemission, Gall \textit{et al.} 
\cite{Gall01} concluded that ScN is a semiconductor with an indirect 
$\Gamma-X$ band gap ($E_g^{\Gamma-X}$) of 1.3$\pm$0.3\,eV. The sizable 
error bar of 0.3 eV has
been mainly attributed to the large background carrier concentration 
of $\sim 5\times 10^{20}$\,cm$^{-3}$ causing an apparent increase of
the band gap due to the Burnstein-Moss shift \cite{mossBurstein}. Reducing 
the electron carrier concentration to 4.8$\times 10^{18}$\,cm$^{-3}$ and
combining tunneling spectroscopy and optical absorption measurements,
Al-Brithen \textit{et al.}  \cite{Smith04} were able to reduce the 
error bar and found a value for $E_g^{\Gamma-X}$ of 0.9$\pm$0.1\,eV. 

Early Kohn-Sham density functional theory (KS-DFT) calculations employing 
the local-density (LDA) or X$\alpha$ approximations predicted ScN to be 
a semimetal with a small negative band gap between \mbox{-0.01} and 
\mbox{-0.21\,eV} 
\cite{Monnier85,Neckel76,Eibler83}. In order to overcome the well known 
underestimation of the LDA band gap, more advanced exact-exchange 
(OEPx(cLDA)) \cite{Gall01} and screened exchange \cite{Stampfl01} 
calculations have been performed, and showed 
that ScN is a semiconductor with an indirect $\Gamma$ to $X$ band gap, in accord 
with experimental evidence \cite{Smith04,Gall01}. However, the calculated band gap of 
1.60\,eV found in both studies is significantly larger than the most recent 
experimental value of 0.9$\pm$0.1\,eV \cite{Smith04}. 

In order to shed light on this discrepancy we have performed quasiparticle energy
calculations in Hedin's $GW$ approximation \cite{Hedin65},
which is a well established technique to calculate accurate band structure energies and
currently the choice for computing quasiparticle band structures of solids
\cite{Onida/Reining/Rubio:2002,Aulbur-rev,Patrick05}.
The quasiparticle calculations predict ScN in the rock-salt phase to 
have an indirect band gap between the $\Gamma$ and $X$ point
of  0.99$\pm$0.15\,eV, strongly supporting recent experimental findings.
In addition we have also determined  the direct band gaps and other 
electronic structure parameters relevant for device simulations:
the volume deformation potentials of the main band gaps and the 
longitudinal and transverse effective masses of the conduction band
at the $X$ point. The effective mass has previously  been calculated at the 
level of the LDA \cite{meff_LDA}, but 
to the best of our knowledge 
only one experimental study has reported a conduction band
effective mass for ScN (between 0.1 and 0.2 $m_0$) \cite{Harbeke} so far.
For the deformation potentials no 
experimental or theoretical data is available, yet.

Most commonly, the Greens function $G_0$ and the screened
potential $W_0$ required in the $GW$ approach (henceforth denoted $G_0W_0$) 
are calculated from a set of KS-DFT 
single particle energies and wave functions 
$\{\epsilon_i,\phi_i\}$. 
Since $G_0$ and $W_0$ are not usually updated by the 
quasiparticle wave functions and energies $\{\epsilon_i^{QP},\phi_i^{QP}\}$ 
in a self-consistent manner, the quasiparticle energies 
depend on the approximation used to calculate the input data
\cite{Li02,Li05,Patrick05,Patrick06}. 

Originally, $G_0W_0$ calculations were based on LDA
data (LDA-$G_0W_0$) and were found to accurately predict  
band gaps of $sp$-bonded semiconductors (with a typical error bar of $\sim$0.1\,eV)
\cite{Aulbur-rev}. 
However, complications arise when the LDA-$G_0W_0$ approach is used to calculate the 
electronic structure of 
semiconductors with \textit{negative} LDA band gaps \cite{Kotani02,Usuda04} 
or when occupied shallow semicore $d$ bands are treated as valence 
in the pseudopotential framework \cite{GWCdS,Rohlfing98,Luo/Louie:2002}. For 
such semiconductors, $G_0W_0$ calculations based on OEPx(cLDA) data (OEPx(cLDA)-$G_0W_0$) have been 
found to provide a reliable tool to obtain band gaps with an accuracy 
of 0.1 eV \cite{Patrick05,Patrick06}.

The key to the improved description in the OEPx(cLDA)-$G_0W_0$ approach
can be found in the treatment of exchange. 
In the exact-exchange KS approach the formal expression for the total energy is the
same as in Hartree-Fock. The difference between the two methods lies in the potential felt by the
electrons: in Hartree-Fock the exchange potential is non-local, whereas in the exact-exchange KS
approach it is local and constructed to be the variationally best local ground state potential to
the non-local Hartree-Fock exchange potential \cite{Casida:1995}. 
Like Hartree-Fock, the exact-exchange KS approach is therefore free of
self-interaction, but since the eigenvalues are solutions to a local potential
they are in general closer to (inverse) photoemission data for semiconductors 
than the Hartree-Fock single particle energies 
\cite{Staedele97,Staedele99,Aulbur}.
The $GW$ formalism, on the other hand, goes beyond the KS approach and describes
the interaction of weakly correlated quasiparticles by means of a non-local,
energy dependent self-energy. It takes the
form of the non-local exchange potential encountered in the Hartree-Fock
approach, which is screened by correlation in the random-phase approximation
(RPA).
To elucidate the effects of exchange and correlation on the electronic 
structure of ScN we therefore first analyze the influence of exchange by 
comparing LDA (GGA) and OEPx(cLDA) calculations, before turning to the
difference between OEPx(cLDA) and $G_0W_0$.

The paper is organized as follows. Sec.~\ref{sec:CM} describes our 
computational approach. The results are presented and discussed in Sec.~\ref{sec:conc}. 
Finally,  a summary is given in Sec.~\ref{sec:conc}. 

%%%%%%%%%%%%%%%%%%%%%%%%%%%%%%%%%%%%%%%%%%%%%%%%%%%%%%%%%%%%%%%%%%%%%%%
\section{Computational Method}
\label{sec:CM}

The KS-DFT calculations have been performed with the 
{\em ab initio} pseudopotential (PP) plane-wave code \texttt{SPHIngX}  \cite{sphingx}.
A consistent set of norm-conserving scalar-relativistic PPs has been used
for each of the exchange-correlation functionals (LDA, GGA and OEPx(cLDA)). 
The OEPx(cLDA)-PPs have been constructed as described in Ref. \onlinecite{EXXPP} and the LDA 
and GGA ones using the \texttt{FHI98PP} code \cite{Fuchs}.
All PPs for both Sc and N have been generated according to the Troullier-Martins 
optimization scheme \cite{TM} and have then been transformed into the separable 
Kleinman-Bylander form \cite{KB}.
% rewritten 
Unlike in the group-IIIA-nitrides GaN and InN 
the cation 3$s$ and 3$p$ states have moderate binding energies in ScN and 
the Sc 3$p$ and to a lesser degree also the Sc 3$s$ electrons couple to the upper 
valence bands as a partial charge density analysis reveals (see Section \ref{sec:BS} and 
also Fig. 4 in Ref. \onlinecite{Stampfl01}).
For Sc, the entire semicore shell (3$s$, 3$p$ and 3$d$ states) has therefore been treated 
as valence with an ionic configuration [Ne]3$s^2$3$p^6$3$d^1$. 
% end rewritten
Following Ref. \onlinecite{Stampfl99}, only the 2$s$ and 2$p$ components have been included for N. 
Adding a $d$ component for N yields negligible effects on the calculated band 
structure and total energies of ScN. For Sc, we have chosen a core radius of 1.4, 
1.4, and 1.8 bohr for the $s$, $p$ and $d$ orbitals, respectively. For N, 
a common core radius of 1.5 bohr has been adopted. 
The $s$ ($p$) component is taken as the local component for Sc (N). 
Only one projector per angular momentum channel has been used, 
i.e. 3$p$ and 3$d$ for Sc. We have verified that this procedure does not
compromise higher lying states in the same channel by ensuring
that the eigenvalues of the 4$s$ and 4$p$ levels in the pseudo-atom reproduce 
those of the all-electron calculation. % to within 0.05 eV.
These pseudopotentials have been carefully 
tested (see also below) and are free of ghost states \cite{GhostStat}. 

For the LDA calculations we have used the Ceperley-Alder \cite{CA} exchange-correlation 
data as parametrized by Perdew and Zunger \cite{PZ}. The GGA 
calculations have been performed with the Perdew-Burke-Ernzerhof \cite{PBE} 
functional. In the OEPx(cLDA) calculations, the exchange energy and potential have been treated 
exactly and correlation has been added on the LDA level. 
Throughout the paper, the combination of exact-exchange and LDA 
correlation will be referred to as OEPx(cLDA).

The $G_0W_0$ calculations have been performed employing the $GW$ space-time approach 
\cite{STGW}, in the \texttt{gwst} implementation \cite{gwst1,gwst2,gwst3}. 
The Kohn-Sham 
eigenvalues and wavefunctions ($\epsilon_i$ and $\phi_i$) in either
OEPx(cLDA) or LDA are used as input to construct $G_0$ and $W_0$. Head and wings 
of the dielectric matrix (which converge slowly with respect to the {\bf k}-mesh) 
have been calculated using a fine $10\times10\times10$ Monkhorst-Pack (MP) 
mesh \cite{MP}. We find that using an offset of 
[$\frac{1}{2}$,$\frac{1}{2}$,$\frac{1}{2}$] yields faster
convergence with respect to the number of $k$-points, because the $k$-point 
set then contains fewer high-symmetry points. Contributions arising from 
the non-local part of the pseudopotential are fully taken into account
\cite{gwst3}. A regular 
$4\times4\times4$ MP mesh centered on the $\Gamma$-point 
then proves to be sufficient for the full $G_0W_0$ calculations. 

Brillouin zone integrations in the DFT calculations have been performed on a 
$4\times4\times4$ MP mesh. In all calculations an 
energy cutoff of 80 Ryd is used for the plane wave expansion of the 
wavefunctions. 
For the independent particle polarizability $\chi_0$, 
which enters in the calculation of the OEPx(cLDA) potential (see e.g.
Ref.~\onlinecite{Staedele99}), an energy cutoff of 55 Ryd gives 
converged results. Conduction band states up to the same energy cutoff have been 
included in the calculation of the electronic Green's
function in the OEPx(cLDA) as well as in the $GW$ calculations. 
These parameters yield converged KS and quasiparticle energies to within 
0.05 eV. 

\begin{table}
  \begin{ruledtabular}
    \begin{tabular}{lddd}
    \multicolumn{1}{l}{Approach}   & 
    \multicolumn{1}{c}{ $a_0$ (\AA)}  &
    \multicolumn{1}{c}{   $B_0$ (GPa)} & 
    \multicolumn{1}{c}{$B_0^{\prime}$} \\
    \hline
    \multicolumn{4}{l}{Present work} \\ 
    \hline
    PP-PW(LDA)    &  4.455    &   221  & 4.27 \\
    PP-PW(GGA)    &  4.533    &   196  & 4.36 \\
%    PP-PW(EXX)    &  4.461    &   258  & 4.00 \\ 
    \hline
    \multicolumn{4}{l}{Other theoretical calculations} \\ 
    \hline
    FP-LAPW(LDA) \cite{Stampfl01}     &  4.42     &   235  & \\ 
    FP-LAPW(LDA) \cite{Takeuchi02}    &  4.44     &   220  & \\ 
    FP-LAPW(GGA) \cite{Stampfl01}     &  4.50     &   201  & \\ 
    FP-LAPW(GGA) \cite{Takeuchi02}    &  4.54     &   201  & 3.31 \\ 
    \hline
    Experiment \cite{Gall98}          &  4.501    &  
      \multicolumn{1}{c}{182$\pm$40} & \\
    \end{tabular}
  \end{ruledtabular}
  \caption{\label{tab:structure} Structural parameters of ScN: 
           lattice constant ($a_0$), bulk modulus ($B_0$) 
	   and its derivative ($B_0^{\prime}$) calculated in our pseudopotential,
	   plane-wave (PP-PW) approach  compared to previous results obtained 
	   with the all-electron full-potential linearised augmented plane-wave 
	   approach (FP-LAPW) and experiment.}
\end{table}

We have tested the reliability of our pseudopotential calculations against
all-electron calculations in the full potential linearized augmented plane wave
(FP-LAPW) approach, where available.
For ground state calculations on the level of LDA and GGA 
we find good agreement for the structural properties, as 
Table~\ref{tab:structure} illustrates.
LDA slightly underestimates the lattice constant of ScN compared to
experiment \cite{Gall98} by approximately 1\%, whereas GGA results in a slight
overestimation.

The issue of performing self-consistent $GW$ calculations is still a matter of 
debate \cite{Ku/Eguiluz:2002,Tiago/Ismail-Beigi/Louie:2003,
             Delaney/Garcia-Gonzalez/Rubio/Rinke/Godby:2004,
             Zein/Savrasov/Kotliar:2006,Schilfgaarde/Kotani/Faleev:2006}.
Unlike in DFT, a self-consistent solution of the full set of equations 
for the self-energy in many-body perturbation theory 
would go beyond the $GW$ approximation and 
successively introduce higher order electron-electron 
interactions with every iteration step.
Solving the $GW$ equations self-consistently is therefore inconsistent
if no higher order electron-electron interactions are included. 
It was first observed for the homogeneous electron 
gas \cite{Holm/vonBarth:1998} that the spectral features broaden
with increasing number of iterations in the self-consistency cycle. 
Similarly, for closed shell atoms the good 
agreement with experiment for the ionization energy after the first
iteration is lost upon iterating the equations to self-consistency
\cite{Delaney/Garcia-Gonzalez/Rubio/Rinke/Godby:2004}. 
Imposing self-consistency in an approximate fashion
\cite{Luo/Louie:2002,Fleszar/Hanke:2005,
      Marsili/etal:2005,Schilfgaarde/Kotani/Faleev:2006}
is not unique and different methods yield different results.
Since the issue of self-consistency within $GW$ is still discussed controversially,
we refrain from any self-consistent treatment and remain with the zeroth order in the 
self-energy ($G_0W_0$). We argue (see Section \ref{sec:BS}) that in the case of ScN 
the error bar resulting from this approximation is only of the order of 0.15 eV.

%%%%%%%%%%%%%%%%%%%%%%%%%%%%%%%%%%%%%%%%%%%%%%%%%%%%%%%%%%%%%%%%%
\section{Results and discussion}
\label{sec:results}

\subsection{Electronic band structure}
\label{sec:BS}

\begin{table} 
  \begin{ruledtabular}
    \begin{tabular}{lddd}
      \multicolumn{1}{l}{Approach}  &   
      \multicolumn{1}{c}{$E_g^{{\Gamma}-{\Gamma}}$} &   
      \multicolumn{1}{c}{$E_g^{{\Gamma}-X}$} & 
      \multicolumn{1}{c}{$E_g^{X-X}$} \\
      \hline
      \multicolumn{4}{l}{Present work} \\ 
      \hline
      OEPx(cLDA)-$G_0W_0$     & 3.51   &  0.84  &   1.98  \\
      LDA-$G_0W_0$     & 3.71   &  1.14  &   2.06  \\
      $[G_0W_0]_{\rm average}$ & 3.62   &  0.99  & 2.02 \\
      OEPx(cLDA)              & 4.53   &  1.70  &   2.59  \\
      GGA              & 2.43   & -0.03  &   0.87  \\
      LDA              & 2.34   & -0.15  &   0.75  \\
      \hline
      \multicolumn{4}{l}{Other theoretical work}  \\
      \hline
      OEPx(cLDA)\cite{Gall01}     &  4.70 &    1.60     & 2.90  \\
      sX\cite{Stampfl01}   &       &    1.58     & 2.41  \\
      \hline
      \multicolumn{4}{l}{Experiment}  \\
      \hline
      Ref. \onlinecite{Gall01}    &  
	 \multicolumn{1}{c}{$\sim$3.8}  &  1.30        & 2.40  \\
      Ref. \onlinecite{Smith04}   &             
                                  &  \multicolumn{1}{c}{0.9$\pm$0.1} &  2.15 \\
    \end{tabular}
  \end{ruledtabular}
  \caption{\label{tab:gaps} Calculated and experimental band gaps ($E_g$)
           of ScN (in eV). sX denotes previous screened exchange calculations 
	   and $[G_0W_0]_{\rm average}$ the arithmetic average between 
	   the OEPx(cLDA)-$G_0W_0$ and LDA-$G_0W_0$ results (see text).}
\end{table}

\begin{figure} %90 99 528 547
  \epsfig{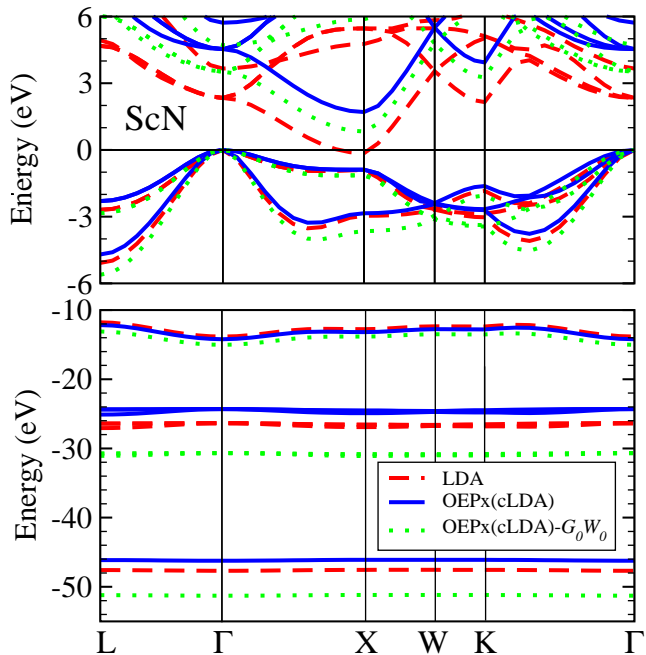}
  \caption{\label{fig:BST} Electronic band structure of rock-salt ScN in LDA,
           OEPx(cLDA) and OEPx(cLDA)-$G_0W_0$. The top panel shows the upper 
	   valence and lower conduction bands, aligned at the top of the 
	   valence band. The bands in the lower panel have mainly N 2$s$ 
	   (around -13~eV),  Sc 3$p$ (around -30~eV) and Sc 3$s$ 
	   character (around -50~eV). }
\end{figure}

The quasiparticle band structure of ScN is calculated employing both the OEPx(cLDA)-$G_0W_0$ and 
LDA-$G_0W_0$ approach. To understand the effect of the starting point on the $G_0W_0$ 
calculations we first analyze the KS band structure
using three levels of approximations for the XC potential (OEPx(cLDA), GGA and LDA). 
For a meaningful comparison between the results of these calculations among themselves 
and with experiment, these electronic structure calculations have been performed 
at the experimental equilibrium volume. As an example, we show in Fig.~\ref{fig:BST} 
the electronic band structures of ScN in LDA, OEPx(cLDA) and OEPx(cLDA)-$G_0W_0$.
Table \ref{tab:gaps} summarizes the 
calculated band gaps ($E_g^{\Gamma-X}$, $E_g^{X-X}$ and $E_g^{\Gamma-\Gamma}$), 
previous OEPx(cLDA) \cite{Gall01} and screened exchange 
\cite{Stampfl01} results  and experimental data \cite{Smith04,Gall01}. For the 
following discussion we consider only the latest experimental data of 
Al-Brithen \textit{et al.} on low background carrier samples \cite{Smith04} as a 
reference.

\begin{figure} % 22 106 596 473
    \epsfig{bbllx=22,bblly=106,bburx=596,bbury=473,clip=,
            file=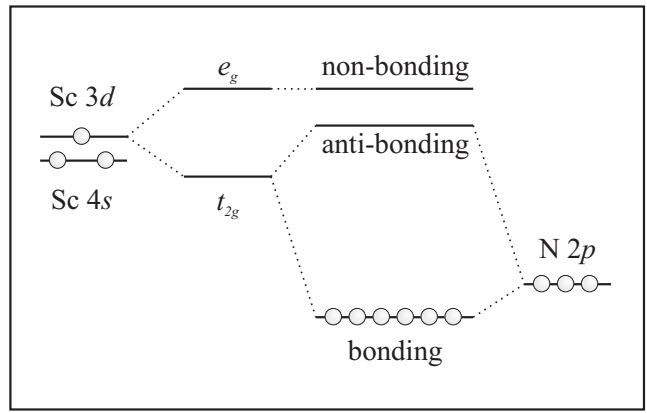,width=0.98\columnwidth}
  \caption{\label{fig:bonding} 
           Schematic diagram of the $pd$ bonding in rock-salt ScN: the Sc 3$d$ states are split
	   into $t_{2g}$ and $e_g$ states by the crystal field. The $t_{2g}$ states 
	   then form bonding
	   and anti-bonding bands with the N 2$p$ states and the $e_g$ states 
	   non-bonding bands that lie higher in energy.
	   The occupation of the relevant states
           and bands is shown by circles.}  
\end{figure}

% Bonding ....  
In ScN the scandium atom donates its two 4$s$ and single 3$d$ electron to the
nitrogen atom.
According to the bonding analysis of Harrison and Straub \cite{Harrison}, the 
five $d$ states of Sc hybridize with the three valence $p$ states of the 
neighboring N atoms in the rock-salt structure of ScN, 
forming three $p$-like bonding, three $d$-like anti-bonding $t_{2g}$ 
and two $d$-like non-bonding $e_g$ bands. 
The bonding scheme together with the electron filling 
of these bands is sketched in Fig.~\ref{fig:bonding}. 
Performing a partial charge density analysis we have confirmed that the upper three 
valence bands in the DFT calculations correspond to the bonding states and 
originate mainly from the 
N 2$p$ states with some admixture of the Sc 3$d$ states, while the lowest 
conduction band are the anti-bonding $t_{2g}$ states with Sc 3$d$ character. The two
bands derived from the non-bonding $e_g$ states are around 1.2~eV higher in energy.
This assignment is consistent with the partial density of states analysis of 
Stampfl \textit{et al.} \cite{Stampfl01}. The character of the deeper lying
bands is given in the caption of Fig.~\ref{fig:BST}.

% LDA, GGA and OEPx(cLDA) band gaps
Focusing first on the band gaps presented in Table \ref{tab:gaps} we note 
that both LDA and GGA underestimate all band gaps by more than 100\%. 
The GGA band gaps are only marginally ($\sim$ 0.1 eV) larger than those of LDA.
The reason is a combination of three factors: 1) in direct and inverse photoemission
experiments electron addition and removal energies are probed, but the derivative 
discontinuity of the exchange-correlation potential with respect to changes in the 
particle number is not taken into account in KS-DFT single-particle energy 
calculations \cite{Scham/Schlueter}
2) LDA and GGA are approximate exchange-correlation functionals, which 3) suffer from
inherent self-interaction effects.
The OEPx(cLDA) formalism also does not fulfill criterion 2), but it is self-interaction 
free. This leads to a significant opening of the Kohn-Sham band gaps compared to those of 
LDA and GGA \cite{Staedele99,Staedele97,Aulbur} as is evident from Table~\ref{tab:gaps}. 
Although the OEPx(cLDA) formalism exhibits a derivative discontinuity 
 \cite{Krieger92,Casida:1999} and therefore fulfills criterion 1) this is of no benefit
in KS-DFT single-particle energy calculations. 
When the excitation energies are calculated by computing
total energy differences in OEPx(cLDA) between the $N$ and the $N\pm$1 electron 
system (frequently denoted $\Delta$ self-consistent field ($\Delta$SCF) approach),
the derivative discontinuity is taken into account properly \cite{Krieger92}. 
In KS-DFT, however, the excitation energies are approximated by Kohn-Sham eigenvalue
differences of the $N$-electron system alone. The derivative discontinuity does
hence not enter the calculation and all states experience the
same exchange-correlation potential.

\begin{figure}% 157 167 632 449
    \epsfig{bbllx=157,bblly=167,bburx=632,bbury=449,clip=,
            file=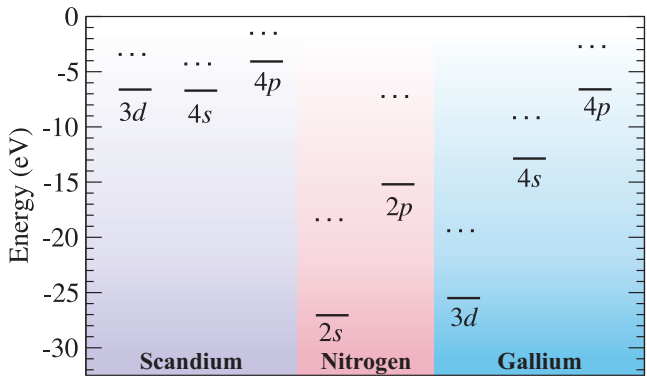,width=0.98\columnwidth}  
  \caption{\label{fig:eigenval} 
         Eigenvalue spectrum of the isolated Sc atom compared to N and Ga. 
         Dotted lines show the levels in the LDA and solid lines in OEPx(cLDA). }  
\end{figure}

\begin{figure}%20 20 423 822
  \epsfig{bbllx=20,bblly=20,bburx=423,bbury=822,clip=,
          file=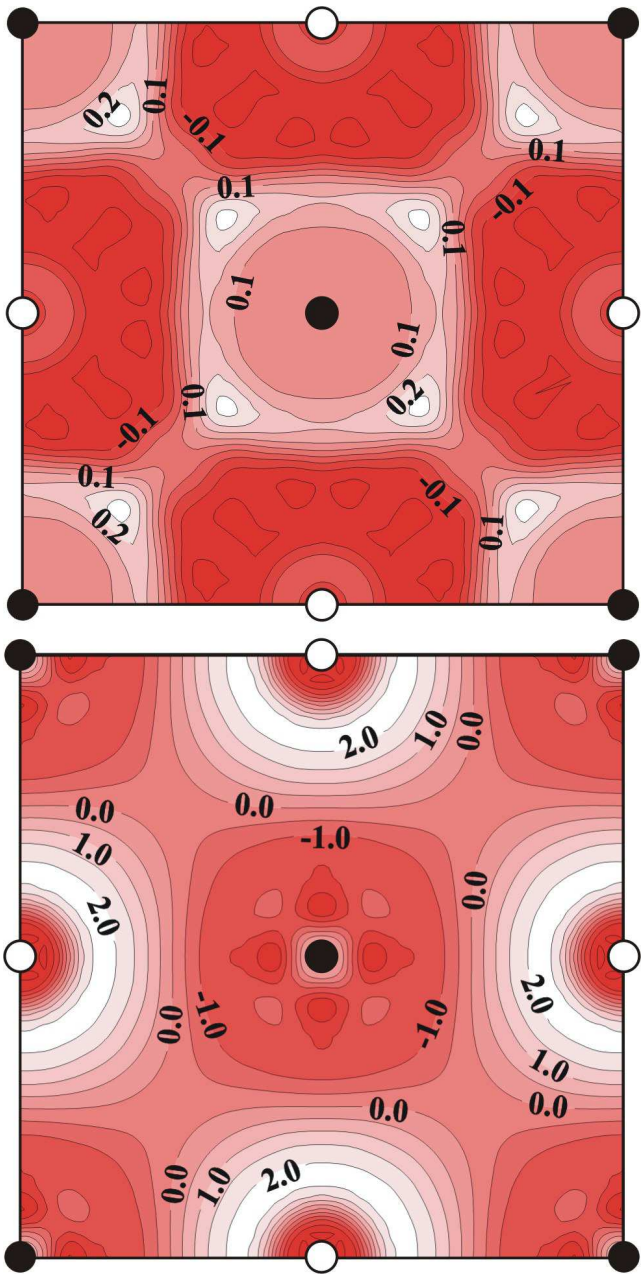,width=0.98\columnwidth}
  \caption{\label{fig:rho} The difference between (a) the OEPx(cLDA) and LDA exchange 
           potentials (in Hartree) and (b) the electronic charge densities 
          (in electrons/unit cell) of ScN for one of the square faces of 
	  the conventional rock-salt unit cell. Black circles denote Sc and
	  white circles N atoms.} 
\end{figure}

% Opening of Eg 
Having established that the removal of the self-interaction in the OEPx(cLDA)-KS 
approach is the distinguishing feature compared to KS-LDA or KS-GGA calculations
we will now  illustrate how this leads to an opening of the band gap in ScN. 
For this it is illuminating to start from 
the eigenvalues of the isolated Sc and N atoms, depicted in 
Fig.~\ref{fig:eigenval}. 
The removal of the self-interaction in OEPx(cLDA) leads to a downward shift of all
atomic states. Since the electrons in the 2nd shell of the nitrogen atom 
are more localized than the electrons populating the 3rd and 4th shell in scandium the
self-interaction correction to the N 2$p$ state is much larger than that of the Sc 3$d$
state.
Inspection of the difference between the exchange potential in OEPx(cLDA) ($v_x^{\rm OEPx(cLDA)}$)
and LDA ($v_x^{\rm LDA}$) shown in Fig.~\ref{fig:rho}(a) 
reveals that the large relative shift of the atomic N 2$p$ state also translates to 
the solid. Fig.~\ref{fig:rho}(a) illustrates that $v_x^{\rm OEPx(cLDA)}$ is 
significantly higher than $v_x^{\rm LDA}$ in the Sc regions and lower
around the N atoms. This difference in $v_x$ leads to a significant charge density 
redistribution [shown in Fig.~\ref{fig:rho}(b)].  The charge transfer from the Sc to 
the N regions gives rise to an increase in the bond ionicity, which, in turn,
leads to an opening of the band gap --- consistent with our OEPx(cLDA) band structure 
calculations.

In the II-VI compounds and group-III-nitrides this mechanism is also responsible for an opening
of the band gap in OEPx(cLDA) compared to LDA, but it is complemented
by a contribution arising from the coupling between the anion semicore $d$ electrons and 
the 2$p$ electrons of nitrogen.
Taking GaN as example again the Ga 3$d$ electrons of gallium are energetically lower 
than the 2$p$ electrons of nitrogen, while in Scandium the Sc 3$d$ lie above the N 2$p$ states
(cf Fig.~\ref{fig:eigenval}). Not the anion 3$d$, like in ScN, but the Ga 4$s$ electrons thus
form the lower conduction bands with the nitrogen 2$p$ states.
Since the OEPx(cLDA) shift of the N 2$p$ states is larger than that of the Ga 4$s$ states the
bond ionicity and hence the band gap increase just like in ScN. 
In addition the Ga 3$d$ electrons localize stronger in GaN when the self-interaction is 
removed in the OEPx(cLDA) approach \cite{Patrick05}. 
As a result the $pd$ repulsion reduces and the valence bands are 
lowered in energy leading to a further opening of the band gap \cite{Qteish05}.

% LDA+GW band gaps
We now turn to the quasiparticle band structure. In the OEPx(cLDA)-$G_0W_0$ approach the band
structure is calculated directly at the experimental lattice constant. 
However, the \textit{negative} LDA band gap (see Table \ref{tab:gaps}) impedes the 
application of this direct approach in the LDA-$G_0W_0$ formalism 
with the \texttt{gwst} code, 
since in its current implementation\cite{gwst1,gwst2} a clear separation between 
conduction and valence bands is required. 
Therefore, an indirect approach is adopted. First, LDA-$G_0W_0$ 
calculations are performed at a lattice constant ($a_0$ = 4.75 \AA) larger than 
the experimental one, where the fundamental band gap in the LDA is small but positive. 
We then use the LDA \textit{volume} deformation potentials (see Section
\ref{sec:meff_defpot}) to determine the 
corresponding LDA-$G_0W_0$ band gaps at the equilibrium lattice constant.
Using the volume deformation potentials of the LDA instead of the  
quasiparticle ones is a well justified approximation, as we will show in the next 
subsection. 
While this approach is in principle not limited to band gaps, it proves to be too
cumbersome for a whole band structure calculation, because for every band structure
point the corresponding deformation potential would have to be determined.

% OEPx(cLDA)+GW band gaps
The quasiparticle band structure is shown in Fig.~\ref{fig:BST} and the direct 
band gaps at the $\Gamma$ and X-point are presented together with the indirect 
gap between $\Gamma$ and X in Table \ref{tab:gaps}. 
It is interesting to note that the LDA-$G_0W_0$ and OEPx(cLDA)-$G_0W_0$ 
calculations, starting from the two extremes (negative 
band gap in LDA, 0.8 eV overestimation in OEPx(cLDA)), yield quasiparticle band gaps that agree to
within 0.3\,eV.
Since the LDA-based calculations are close to the limit of metallic screening,
whereas the OEPx(cLDA)-based calculations form the opposite extreme of starting from 
a completely self-interaction free exchange-correlation functional,
we expect the results of a self-consistent $GW$ calculation to fall in the range
between the LDA-$G_0W_0$ and OEPx(cLDA)-$G_0W_0$ calculations. 
From these results we estimate the error bar associated with omitting self-consistency in
$GW$ to be of the order of 0.15 eV for ScN.
Taking OEPx(cLDA)-$G_0W_0$ results as lower and those of the LDA-$G_0W_0$ as upper 
bounds, the arithmetic averages for $E_g^{\Gamma-X}$, $E_g^{{\Gamma}-{\Gamma}}$ 
and $E_g^{X-X}$ are 0.99, 3.62 and 2.02\,eV, respectively. These are significantly
lower than those from OEPx(cLDA) and the more approximate screened exchange 
calculations, as Tab.~\ref{tab:gaps} demonstrates. Our quasiparticle gaps 
clearly support recent experimental findings of an indirect gap 
of 0.9$\pm$0.1\,eV \cite{Smith04} and 
are at the lower bound of earlier measurements on samples with 
unintentionally high background carrier concentration (1.3$\pm$0.3\,eV) \cite{Gall01}.

% atomic all-electron eigenvalues in LDA (in eV)
%                                         Ga                   Sc
%  1     1     0        2.0000       -10210.7601          -4385.1827
%  2     2     0        2.0000        -1260.0689           -472.9569
%  3     2     1        6.0000        -1098.2169           -388.0270
%  4     3     0        2.0000         -147.4363            -54.7142
%  5     3     1        6.0000          -98.6601            -33.5675
%  6     3     2       10.0000          -19.4072             -3.4262
%  7     4     0        2.0000           -9.1703             -4.2925
%  8     4     1        1.0000           -2.7331              

The fact that LDA-$G_0W_0$ and OEPx(cLDA)-$G_0W_0$ calculations yield very 
similar quasiparticle band gaps is in disagreement with our previous
observation for II-VI compounds and GaN \cite{Patrick05}.
The difference between ScN and these compounds is that for the latter 
the cation semicore $d$ shell is fully filled and the remaining $s$ and $p$ electrons 
in the semicore shell are much lower in energy. Taking GaN as an example the 3$s$ 
electrons in the Gallium atom are approximately 100\,eV  and the
3$p$ electrons approximately 60\,eV lower than in Scandium. Unlike in ScN the 3$p$
derived bands therefore show no noticeable dispersion in GaN (cf Fig.~\ref{fig:BST}).
Resolving these more strongly localized 3$s$ and 3$p$ electrons in GaN with plane-waves
will thus require significantly higher plane-wave cutoffs \cite{Patrick05} than the 
80~Ry used in the present study for ScN.
In a pseudopotential framework it would
hence make sense to explicitly include the $d$ electrons of the cations in the II-VI
compounds and group-III-nitrides as valence electrons,
but to freeze the chemically inert semicore $s$ and $p$ electrons in the core of the 
pseudopotential.
However, due to the large spatial overlap of the atomic semicore $s$ and $p$ with the
$d$ wavefunctions, core-valence exchange is
large in these compounds. As a consequence core-valence exchange is treated 
inconsistently when going from LDA to LDA-$G_0W_0$, if pseudopotentials are
used in this fashion, because the exchange self-energy in the $GW$ approach acts
on the $d$ electrons in the solid, but cannot act on the $s$ and $p$ electrons in 
the semicore shell, too. The result is a severe underestimation of the LDA-$G_0W_0$ band 
gaps and $d$-bands that are pushed energetically into the $p$-derived valence 
bands in the II-VI compounds \cite{Patrick05,GWCdS,Rohlfing98}.

The only way to remedy
this problem within LDA-$G_0W_0$ is to free the electrons in question by 
performing all-electron $G_0W_0$ calculations \cite{Kotani02}
or by using pseudopotentials that include the entire shell as valence 
electrons \cite{GWCdS,Rohlfing98,Luo/Louie:2002}, which in the latter case introduces 
formidably high plane-wave cutoffs. 
If, on the other hand, OEPx(cLDA) is
used for the ground state calculation, then the exchange self-energy already
acts on the semicore $s$ and $p$ states in the generation of the pseudopotential.
Since the exchange self-energy can be linearly decomposed into a core and a valence
contribution no non-linear core corrections \cite{NLCC}
 arise in the Hartree-Fock case and they are
expected to be small for OEPx(cLDA) pseudopotentials \cite{Staedele99}.
We take the fact that the quasiparticle band structure in the OEPx(cLDA)-$G_0W_0$ approach
agrees well with (inverse) photoemission data for these materials as indication that
when switching from the local potential in OEPx(cLDA) to the non-local
self-energy in OEPx(cLDA)-$G_0W_0$ core-valence exchange is treated consistently, 
as long as OEPx(cLDA) pseudopotentials are used \cite{Patrick05}.
Since the semicore $s$ and $p$ states are less localized in ScN it is computationally
feasible to include the entire 3rd shell of Sc as
valence in the pseudopotentials (see Section \ref{sec:CM}) and 
thus to conduct a meaningful comparison between the LDA-$G_0W_0$ and 
OEPx(cLDA)-$G_0W_0$ calculations, which enables us 
to assess the error bar with respect to a self-consistent $GW$ treatment.

It remains to be added that in previous studies, where a significant starting point
dependence in LDA-$G_0W_0$ compared to GGA-$G_0W_0$ calculations was noted, 
this was either due to structural effects \cite{Li02} or significant differences in the 
ground state \cite{Li05} introduced when going from LDA to GGA.
Since all calculations in this work were performed at the 
experimental equilibrium volume the KS band structures in LDA and GGA are very similar
(cf. Tab.~\ref{tab:gaps}) and LDA-$G_0W_0$ and GGA-$G_0W_0$ calculations yield 
essential the same result.

%Sc 3s and 3p bands
Finally, Fig.~\ref{fig:BST} illustrates that  the 
Sc 3$p$ and Sc 3$s$ bands are significantly lowered by
the quasiparticle energy calculations.
We have argued recently \cite{Qteish05} that
this lowering can to a large degree be attributed to charge density relaxation effects 
arising from the removal of an electron from these states.
These effects are accounted for in the $G_0W_0$ approach, but not in DFT Kohn-Sham 
single-particle energy calculations and are larger for more localized states such as the 
Sc 3$s$ and 3$p$ bands.

%%%%%%%%%%%%%%%%%%%%%%%%%%%%%%%%%%%%%%%%%%%%%%%%%%%%%%%%%%%%%%%
\subsection{Effective masses and deformation potentials}
\label{sec:meff_defpot}

In this final part we present additional band structure parameters 
of ScN, namely the transverse ($m_t^{*}$) and longitudinal ($m_l^{*}$) 
conduction band effective masses -- at the X point -- and the volume band 
gap deformation potentials, extracted from our quasiparticle energy 
calculations. For comparison we will also discuss the corresponding 
values obtained from LDA, GGA and OEPx(cLDA).

\begin{table} 
  \begin{ruledtabular}
    \begin{tabular}{l|dd|ddd}
    Approach  &  
    \multicolumn{1}{c}{$m_t^{*}$}&   
    \multicolumn{1}{c|}{$m_l^{*}$}&   
    \multicolumn{1}{c}{$a_v^{{\Gamma}-{\Gamma}}$}&   
    \multicolumn{1}{c}{$a_v^{{\Gamma}-X}$} & 
    \multicolumn{1}{c}{$a_v^{X-X}$} \\
    \hline
    OEPx(cLDA)-$G_0W_0$ & 0.189 & 1.483 & 1.54	&  2.02  &   2.04  \\
    OEPx(cLDA)          & 0.253 & 1.450 & 1.07	&  2.06  &   2.21  \\
    GGA                 & 0.139 & 1.625 & 1.43	&  1.87  &   1.92  \\
    LDA                 & 0.126 & 1.570 & 1.36	&  1.95  &   2.03  \\
    LDA \cite{meff_LDA}  & 0.124 & 1.441 & \\
    \end{tabular}
  \end{ruledtabular}
  \caption{\label{tab:meff/defpot} Transverse ($m_t^{*}$) and longitudinal ($m_l^{*}$)
           effective masses of the conduction electrons  
           at the X point (in units of $m_0$) and different 
	   band gap volume deformation potentials ($a_v^\alpha$) (in eV)
	   for ScN.}
\end{table}

The conduction band effective masses at the X point are calculated by fitting 
a quadratic function to the corresponding band structure energies along the 
$\Delta$ ($\Gamma$-X) and Z (X-W) directions for $m_l^{*}$ and $m_t^{*}$, 
respectively. A small $k$-point spacing of 0.01 in units of 
2$\pi/a$ yields converged effective masses, which are listed
in Table~\ref{tab:meff/defpot}. 
Our LDA results are in good agreement with those of Ref. \onlinecite{meff_LDA}. 
As far as an experimental reference is concerned, we are only aware of one study, 
where a conduction band effective mass between  0.1 and 0.2 $m_0$ has been 
reported\cite{Harbeke}. 
Apart from the OEPx(cLDA) results, all our KS-DFT and quasiparticle energy calculations 
give a transverse effective mass in this range, while the longitudinal effective 
mass is approximately one order of magnitude larger.

\begin{figure}%76 92 464 390
  \epsfig{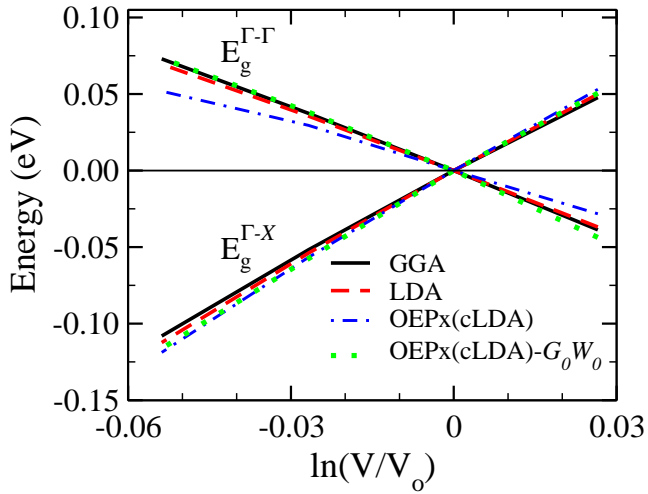}
  \caption{\label{fig:DP} Variation of the indirect band gap ($E_g^{\Gamma-X}$) and 
           the direct band gap at the $\Gamma$-point ($E_g^{\Gamma-\Gamma}$) of ScN 
	   with respect to the corresponding band gaps at the experimental 
	   equilibrium volume ($V_0$), as 
           a function of $\ln (V/V_0)$.} 
\end{figure}

To determine the volume deformation potentials $a_V^{\alpha}$\cite{footnote_defpot} 
for the band gaps $E_g^{\Gamma-\Gamma}$, $E_g^{\Gamma-X}$ and $E_g^{X - X}$, 
the relevant band gaps are 
calculated at four different lattice constants around the experimental equilibrium 
one.
For illustration, we show in Fig.~\ref{fig:DP} the variation of $E_g^{\Gamma-\Gamma}$  
and $E_g^{\Gamma-X}$, with respect to their values at the experimental equilibrium volume,
as a function of $\ln (V/V_0)$.
The corresponding band gaps are then fitted to a quadratic function of 
$\ln V$. The linear deformation potentials obtained in this way are listed in
Table~\ref{tab:meff/defpot}. 
To the best of our knowledge, no experimental or previous theoretical 
results are available for the deformation potentials of ScN.
Based on our results we conclude that $a_V^{\alpha}$ is almost insensitive 
with respect to the computational approaches considered in this article. 
The only exception is $a_V^{\Gamma- \Gamma}$ obtained using the OEPx(cLDA) approach. 
The fact, that the LDA curves in Fig.~\ref{fig:DP} are very close to the OEPx(cLDA)-$G_0W_0$ 
ones, {\it a posteriori} justifies the use of the LDA volume deformation potential
for the indirect calculation of the LDA-$G_0W_0$ band gaps. 

\section{Conclusions }
\label{sec:conc}

Pseudopotential $G_0W_0$ calculations based on Kohn-Sham density-functional theory 
calculations in both the LDA and OEPx(cLDA) have been performed for the electronic structure 
of ScN in the thermodynamically stable rock-salt phase. 
To analyze the effects of exchange and correlation 
the atomic and electronic structure has been studied within DFT for several levels 
of approximations for the exchange-correlation functional (LDA, GGA and OEPx(cLDA)). 
In agreement with previous calculations for ScN, our LDA (OEPx(cLDA))
band gaps are underestimated (overestimated) by about 100\%. 
Despite this large difference, OEPx(cLDA)-$G_0W_0$ and LDA-$G_0W_0$ calculations for the
quasiparticle band structure agree to within 0.3\,eV.
Our quasiparticle gap of 0.99$\pm$0.15\,eV supports the recent observation that ScN has
a much lower indirect band gap than previously thought.
The main advantage of the OEPx(cLDA)-$G_0W_0$ approach lies in the fact that it
facilitates a direct calculation of the 
electronic structure of ScN at the experimental equilibrium
volume, whereas for the LDA-$G_0W_0$ calculation an indirect approach has to be taken 
 due to the negative LDA band gap.

\begin{center}
{\bf {Acknowledgments}}
\end{center}

We acknowledge stimulating discussions with Arthur Smith, Sixten Boeck, Martin 
Fuchs, Matthias Wahn and Christoph Freysoldt.
This work was in part supported by the Volkswagen Stiftung/Germany, the DFG research group 
,,nitride based nanostructures" and
the EU's 6th Framework Programme through the 
NANOQUANTA (NMP4-CT-2004-500198) Network of Excellence.

%references

%%%%%%%%%%%%%%%%%%%%%%%%%%%%%%%%%%%%%%%%%%%%%%%%%%%%%%%%%%%%%%%

%%%%%%%%%%%%%%%%%%%%%%%%%%%%%%%%%%%%%%%%%%%%%%%%%%%%%%%%%%%%%%%
\clearpage

\end{document}